\documentclass[amsmath,amssymb,11pt,superscriptaddress,reprint, preprintnumbers, notitlepage,aps,twocolumn,nofootinbib]{revtex4-1}
\pdfoutput=1 
\usepackage[utf8]{inputenc}
\usepackage[english]{babel}
\usepackage{amsmath}
\usepackage{graphicx}
\usepackage{dcolumn}
\usepackage{pbox}
\usepackage{amssymb}
\usepackage{epsfig}
\usepackage[dvipsnames]{xcolor}
\usepackage[normalem]{ulem}
\usepackage{slashed}
\usepackage{amssymb}
\usepackage{ mathrsfs }
\usepackage{color}
\usepackage[font=small]{caption}
\usepackage[font=small]{subcaption}
\usepackage{url}
\usepackage[makeroom]{cancel}
\usepackage{tikz}
\usetikzlibrary{arrows}
\usepackage{enumitem} 
\usepackage{soul}

\definecolor{MyDarkBlue}{rgb}{0.1, 0.1, 0.8} 
\definecolor{SBlue}{rgb}{0.2, 0.4, 0.7} 
\definecolor{MyLightBlue}{rgb}{0.22,0.51,0.9}
\definecolor{MyGreen}{rgb}{0.0, 0.5, 0.0}
\definecolor{BrickRed}{rgb}{0.8, 0.25, 0.33}
\RequirePackage{hyperref}
\hypersetup{colorlinks, citecolor=SBlue,linkcolor=MyDarkBlue, urlcolor=PineGreen}

\makeatletter
\renewcommand\@makecaption[2]{%
  \par
  \vskip\abovecaptionskip
  \begingroup
  
   \small\rmfamily
    \begingroup
     \samepage
     \flushing
     \let\footnote\@footnotemark@gobble
     \@make@capt@title{#1}{#2}\par
    \endgroup
  \endgroup
  \vskip\belowcaptionskip
}
\makeatother

\makeatletter 
    
\renewcommand\onecolumngrid{
\do@columngrid{one}{\@ne}%
\def\set@footnotewidth{\onecolumngrid}
\def\footnoterule{\kern-6pt\hrule width 1.5in\kern6pt}%
}

\renewcommand\twocolumngrid{
        \def\footnoterule{
        \dimen@\skip\footins\divide\dimen@\thr@@
        \kern-\dimen@\hrule width.5in\kern\dimen@}
        \do@columngrid{mlt}{\tw@}
}%

\makeatother    

\DeclareUnicodeCharacter{2212}{-}
\begin{document}
\title{\vspace{1cm}\Large 
Probing SUSY at Gravitational Wave Observatories
}

\author{\bf Stefan Antusch}
\email[E-mail:]{stefan.antusch@unibas.ch}
\author{\bf Kevin Hinze}
\email[E-mail:]{kevin.hinze@unibas.ch}
\author{\bf Shaikh Saad}
\email[E-mail:]{shaikh.saad@unibas.ch}
\affiliation{Department of Physics, University of Basel, Klingelbergstrasse\ 82, CH-4056 Basel, Switzerland}
\author{\bf Jonathan Steiner}
\email[E-mail:]{j.steiner@thphys.uni-heidelberg.de}
\affiliation{Institut f\"ur theoretische Physik, Universit\"at Heidelberg, Philosophenweg 19, 69120 Heidelberg, Germany}

\begin{abstract}
Under the assumption that the recent pulsar timing array evidence for a stochastic gravitational wave (GW) background at nanohertz frequencies is generated by metastable cosmic strings, we analyze the potential of present and future GW observatories for probing the change of particle degrees of freedom caused, e.g., by a supersymmetric (SUSY) extension of the Standard Model (SM). We find that signs of the characteristic doubling of degrees of freedom predicted by SUSY could be detected at Einstein Telescope and Cosmic Explorer even if the masses of the SUSY partner particles are as high as about $10^4$ TeV, far above the reach of any currently envisioned particle collider. We also discuss the detection prospects for the case that some entropy production, e.g.\ from a late decaying modulus field inducing a temporary matter domination phase in the evolution of the universe, somewhat dilutes the GW spectrum, delaying discovery of the stochastic GW background at LIGO-Virgo-KAGRA. In our analysis we focus on SUSY, but any theory beyond the SM  predicting a significant increase of particle degrees of freedom could be probed this way.
\end{abstract}

\maketitle
\section{Introduction}

The Standard Model (SM) of elementary particles successfully describes a plethora of experimental observations. But it is nevertheless regarded as only an effective theory, that has to be extended 
to a more complete theory, addressing e.g.~the origin of neutrino masses, dark matter, the baryon asymmetry of the universe, the stability of the electroweak scale and finally the quantum nature of gravity. 
One crucial information we are currently lacking is the energy scale at which new physics extends the SM. Future particle colliders like the FCC-hh \cite{FCC:2018vvp} could  discover new particle degrees of freedom with masses up to $\mathcal O(10)$ TeV. 
In this letter we demonstrate that future gravitational wave (GW) observatories have the potential to discover signs of SM extensions at even higher energy scales, by detecting the effects of the extra particle degrees of freedom on the evolution of the universe. 

Supersymmetry (SUSY) may serve as a benchmark for such considerations. It is motivated, among other things, by its ability to resolve or at least ameliorate the hierarchy problem, i.e.\ the instability of the electroweak scale under quantum corrections. Although a lower mass scale $m_\mathrm{S}$ of the predicted superpartner particles, close to the electroweak scale, is favorable by naturalness arguments, there is no upper bound on $m_\mathrm{S}$. 
As we will show, signs of the doubling of particle degrees of freedom characteristic of SUSY could be probed for superpartner masses up to $\mathcal O(10^4)$~TeV, far beyond the reach of any currently envisioned collider.

The condition under which the planned GW observatories can have this intriguing reach is that there exists a predicted stochastic GW background (SGWB) that allows to measure deviations of it caused by the impact of the extra particle degrees of freedom on the evolution of the universe. Interestingly, recent pulsar timing array (PTA) measurements pointing at 
a SGWB at nanohertz frequencies~\cite{Xu:2023wog,Antoniadis:2023ott,NANOGrav:2023gor,Reardon:2023gzh,InternationalPulsarTimingArray:2023mzf} indicate that such a possibility might indeed exist.  
In particular, metastable cosmic strings~\cite{Kibble:1976sj,Vilenkin:1982hm} are among the best-fitting explanations of the PTA data~\cite{NANOGrav:2023hvm}, which has generated some excitement in the particle physics community (see e.g.\ 
\cite{Antusch:2023zjk,Buchmuller:2023aus,Fu:2023mdu,
Lazarides:2023rqf,Ahmed:2023rky,Afzal:2023cyp,Maji:2023fhv,Ahmed:2023pjl,Afzal:2023kqs,King:2023wkm,Ahmed:2024iyd}, and \cite{Lazarides:2023ksx,Yamada:2023thl,
Lazarides:2023bjd,Pallis:2024mip,Kume:2024adn} for other works on unstable cosmic strings).  
The preferred parameters for the metastable cosmic strings from the PTA measurements lead to a prediction of a SGWB in reach of various future GW observatories, such as the Laser Interferometer Space Antenna (LISA)~\cite{Audley:2017drz}, Big Bang Observer (BBO)~\cite{Corbin:2005ny}, DECi hertz Interferometer Gravitational wave Observatory (DECIGO)~\cite{Seto:2001qf}), Einstein Telescope (ET)~\cite{Sathyaprakash:2012jk} and Cosmic Explorer (CE)~\cite{Evans:2016mbw}. If a SGWB from metastable cosmic strings is indeed be confirmed, this opens up the possibility to discover signs of SUSY up to very high $m_\mathrm{S}$.

In our analysis, we first focus on the sensitivity to signs of SUSY assuming metastable cosmic strings with a standard cosmological history, apart from the extra SUSY particle degrees of freedom. We then explore the potential impact of additional late-time entropy production, which could, e.g., be induced by a late-decaying modulus field, leading to a temporary early matter-dominated phase that dilutes the GW spectrum. This dilution may postpone the detection of the SGWB from a metastable cosmic string network in the currently running observatories, i.e.\ at LIGO-Virgo-KAGRA (LVK)~\cite{LIGOScientific:2014pky,VIRGO:2014yos,KAGRA:2018plz}. While our focus is on the implications of SUSY degrees of freedom on the GW spectrum, we emphasize that our results are more general and applicable to any theoretical framework beyond the SM that predicts a substantial increase in particle degrees of freedom.

\section{Metastable cosmic strings and PTA data}\label{sec:PTA}

Topological defects~\cite{Kibble:1976sj,Vilenkin:2000jqa} can be generated  during phase transitions in the early universe. The most prominent examples of topological defects are monopoles and cosmic strings. The former emerge, e.g., when a compact simple group, for example in the context of Grand Unified Theories (GUTs), is spontaneously broken into a subgroup containing an Abelian factor, whereas e.g.~the breaking of an Abelian group leads to the latter. Monopoles must be diluted away by a phase of cosmic inflation, since their presence would overclose the universe. Cosmic strings, on the other hand, rapidly enter a scaling regime and emit GWs (see e.g.~\cite{Hindmarsh:1994re,Auclair:2019wcv}) over a long period in the history of the universe.

Cosmic strings are metastable if these two aforementioned symmetry breaking scales are close by, and the same Abelian factor involved in the monopole formation takes part in the cosmic string production. Metastable strings eventually decay when monopole-antimonopole pairs spontaneously nucleate along the string cores~\cite{Vilenkin:1982hm}. Such decaying strings have the decay rate per string unit length given by~\cite{Vilenkin:1982hm,Preskill:1992ck,Monin:2008mp,Leblond:2009fq,Chitose:2023dam}
\begin{align}
\Gamma_d\simeq \frac{\mu}{2\pi} e^{-\pi\kappa},\;\;\; 
\kappa= \frac{m^2}{\mu}\simeq \frac{8\pi}{g^2} \left( \frac{v_\mathrm{m}}{v_\mathrm{cs}} \right)^2, \label{eq:kappa-Gamma_d}   
\end{align}
where $m\sim 4\pi v_\mathrm{m}/g$ is the mass of the monopole, $\mu\sim 2\pi v_\mathrm{cs}^2$ is the energy per unit length of the string, and $v_\mathrm{m}$ ($v_\mathrm{cs}$) are the vacuum expectation values corresponding to monopole (string) formation. For $\kappa^{1/2}\gg 10$ the network behaves like a stable string network. Due to the decay at time $t_d \simeq \Gamma_d^{-1/2}$ after network formation, metastable cosmic strings have a distinct characteristic GW spectrum at low frequencies, compared to stable cosmic strings.

As mentioned in the introduction, metastable cosmic string networks offer a compelling interpretation of the latest PTA observations. Indeed CPTA~\cite{Xu:2023wog}, EPTA~\cite{Antoniadis:2023ott}, NANOGrav~\cite{NANOGrav:2023gor}, and PPTA~\cite{Reardon:2023gzh} (see also Ref.~\cite{InternationalPulsarTimingArray:2023mzf}) found strong evidence for Hellings-Downs angular correlation~\cite{Hellings:1983fr}, suggesting a SGWB in the nanoherz frequency range. These observations point at a string tension within the range $G\mu \sim 10^{-8} - 10^{-5}$, with $G$ being Newton's constant,  for $\kappa^{1/2} \sim 7.7 - 8.3$, exhibiting a pronounced correlation between the two quantities, as e.g. shown in Fig.~10 of~Ref.~\cite{NANOGrav:2023hvm}. A particularly noteworthy aspect is the overlap of the $68\%$ credible interval  in the $G\mu - \kappa^{1/2}$ parameter space with the third-generation constraints from the LVK collaboration.\footnote{The current LVK bound on the GW amplitude $\Omega_\mathrm{gw}\leqslant 1.7\times 10^{-8}$ at frequency 25 Hz~\cite{KAGRA:2021kbb} leads to the upper limit of $G\mu \lesssim 2\times 10^{-7}$~\cite{LIGOScientific:2021nrg,NANOGrav:2023hvm}. } Moreover, a substantial portion of the $95\%$ credible region aligns well with these constraints, suggesting a preference for $G\mu \lesssim 2\times 10^{-7}$ and $\kappa^{1/2} \sim 8$, if standard cosmological history is assumed.

\begin{figure}[t!]
    \centering
    \includegraphics[width=\linewidth]{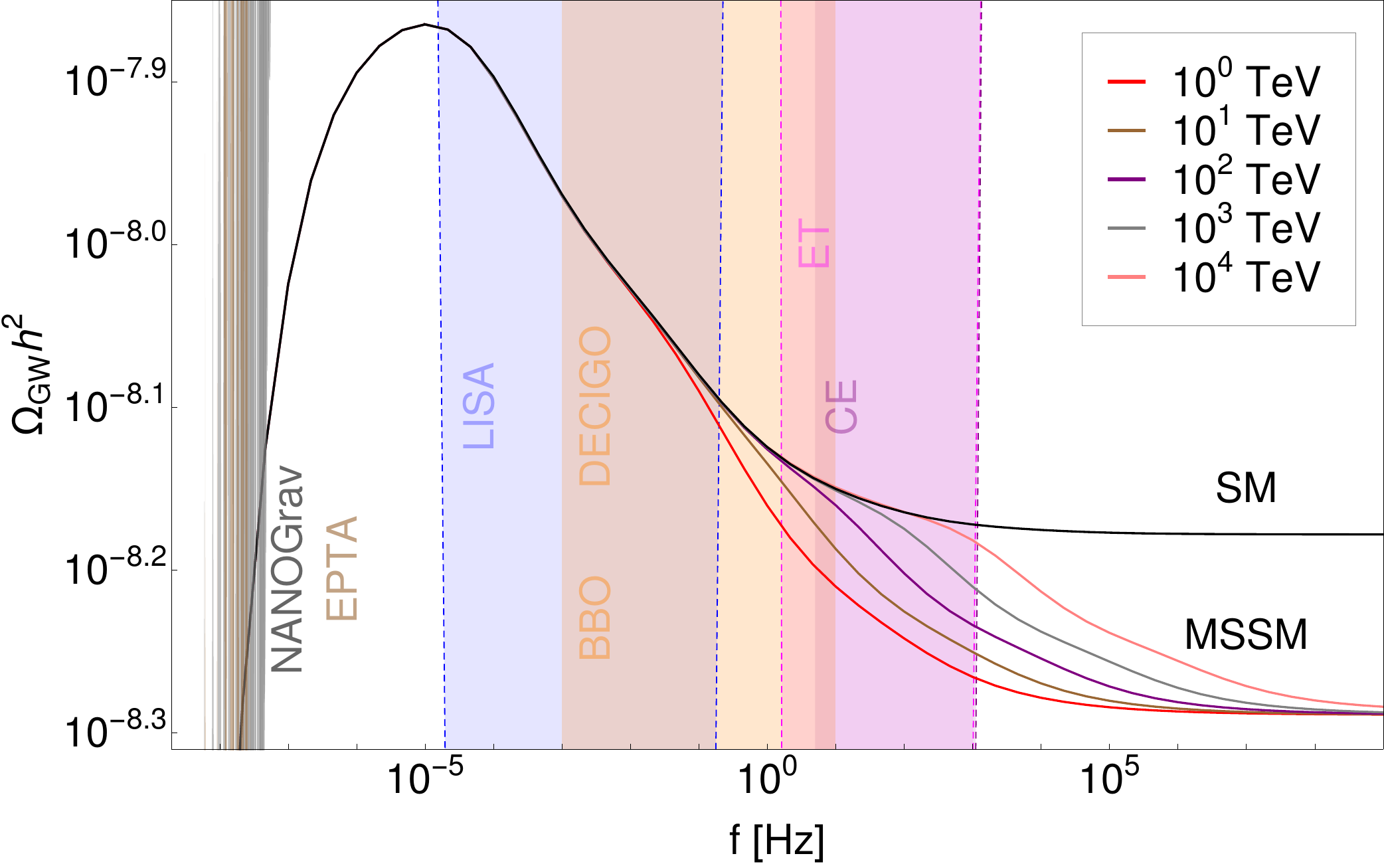}
    \caption{Effect of SUSY degrees of freedom on the GW spectrum of metastable cosmic strings (with $G\mu=1.17 \times 10^{-7}$ and $\kappa^{1/2}=8.115$) for various choices of the SUSY scale $m_\mathrm{S}=10^0,10^1,10^2,10^3,10^4~$TeV. In black we show for comparison the GW spectrum with only the SM degrees of freedom. The power-law integrated sensitivity curves of various future GW observatories are shown together with the recent NANOGrav and EPTA results.}
    \label{fig1}
\end{figure}

Opposed to other sources of GWs, the spectrum of SGWB originating from (metastable) comic strings is relatively flat over a large frequency range (multiple orders of magnitudes). For a radiation dominated universe with a fixed number of particle species above the electroweak scale it forms a calculable plateau at large frequencies, see Fig.~\ref{fig1}. Therefore, it will be possible to fully test whether the observations made by PTAs indeed originate from  metastable cosmic strings in GW observatories, including the currently operating detectors~\cite{LIGOScientific:2014pky,VIRGO:2014yos,KAGRA:2018plz}, as well as upcoming or planned ones~\cite{Audley:2017drz,Corbin:2005ny,Seto:2001qf,Sathyaprakash:2012jk,Evans:2016mbw}. It is this property which enables the search for modifications of the cosmic evolution via deviations from the predicted GW spectrum, such as the ones caused by the extra SUSY particle degrees of freedom.

\section{Effects of SUSY degrees of freedom}\label{sec:dof}
In this section, we focus on the effects of the extra SUSY particle degrees of freedom on the GW spectrum of metastable cosmic strings within the minimal supersymmetric SM (MSSM). To compute the spectrum from a cosmic string network, one needs to determine two basic ingredients: (i) the loop number density, which is the number density $n(\ell, t)$ of sub-horizon sized loops of invariant length $\ell$ at a cosmic time $t$, and (ii) the loop power spectrum, i.e.\ the emitted power $P(f,\ell)$ of GWs of frequency $f$ by a string having length $\ell$. The details of these computations are relegated to Appendix~\ref{sec:apx-a}. Both these factors, and therefore the GW spectrum of a cosmic string network, depend non-trivially on the expansion rate of the universe. This cosmological dependence is encoded in the Hubble rate, which reads 
\begin{align}
H(z)=H_0\left( \Omega_\Lambda + (1+z)^3\Omega_\mathrm{mat} + (1+z)^4 \mathcal{G}(z) \Omega_\mathrm{rad} \right)^{1/2},\label{eq:HCDM}
\end{align}
with $H_0=67.8$ km/s/Mpc, $\Omega_\Lambda=1-\Omega_\mathrm{mat}-\Omega_\mathrm{rad}$, $\Omega_\mathrm{mat}=0.308$, and $\Omega_\mathrm{rad}=9.1476\times 10^{-5}$~\cite{Planck:2018vyg}.

The expansion rate of the universe changes in early times due to the annihilation of relativistic species, which release additional energy into the cosmic plasma, thereby slowing its cooling rate. During the course of this decoupling, entropy is conserved, and the factor that captures the changes in the number of relativistic degrees of freedom is given by
\begin{align}
\mathcal{G}(z)= \frac{g_\ast(z)g^{4/3}_\mathrm{S}(z_0)}{g_\ast(z_0)g^{4/3}_\mathrm{S}(z)}.\label{eq:Gdof}
\end{align}
Here, $g_\ast(z)$ is the effective number of degrees of freedom and $g_\mathrm{S}(z)$ is the effective number of entropic degrees of freedom at redshift $z$ ($z_0$ represents redshift today).

Altering the number of degrees of freedom results in smooth variations in the GW spectrum at frequencies that match the temperature at which this modification occurs. The larger the number of degrees of freedom annihilating at a particular temperature, the more prominent the variations in the spectrum. Assuming all SUSY particles have a common mass scale $m_\mathrm{S}$ with corresponding temperature $T_\mathrm{S}$ (cf.~Appendix~\ref{sec:apx-a} for details), one expects a drastic decrease of the number of degrees of freedom by a factor of $\Delta g_\ast\sim\mathcal{O}(100)$, leaving an identifiable imprint on the GW spectrum~\cite{Battye:1997ji,Cui:2018rwi,Auclair:2019wcv}. The corresponding effect starts to show up at a frequency, $f_\mathrm{S}$, given by~\cite{Cui:2018rwi,Auclair:2019wcv}
\begin{align}\label{eq:fS}
f_\mathrm{S}&\sim (2.1\times 10^{-9}\;\textrm{Hz})\; \left(\frac{m_\mathrm{S}}{\mathrm{GeV}}\right) \left(\alpha \; \Gamma \; G\mu\right)^{-1/2} \\&\times \left(g^\mathrm{SM}_\ast(T_\mathrm{S})+\Delta g_\ast\right)^{5/2}
\left( g^\mathrm{SM}_\ast(T_\mathrm{S}) \right)^{-8/6} \left( g^\mathrm{SM}_\mathrm{S}(T_\mathrm{S}) \right)^{-7/6}, \nonumber    
\end{align}
where $g^\mathrm{SM}_\ast(T_\mathrm{S})=g^\mathrm{SM}_\mathrm{S}(T_\mathrm{S})$ is used, and temperature (frequency) is given in the unit of GeV~(Hz). Moreover, the loop size parameter (defined as $\alpha= t/\ell$) is $\alpha=0.1$, and  the total GW power radiated by each loop is $\Gamma\sim 50$ (cf.\ \cite{Blanco-Pillado:2017oxo}). 

Furthermore, at very high frequencies, the spectrum from a cosmic string network  shows a characteristic flat plateau, which deep in the radiation era is given by~\cite{Blanco-Pillado:2017oxo} 
\begin{align}
\Omega^\mathrm{plateau}_\mathrm{GW}\sim 8\; \Omega_{\mathrm{rad}} \;\mathcal{G}(z) \left( \Gamma^{-1} G\mu \right)^{1/2}.   \label{eq:plateau}
\end{align}
The decoupling of the additional degrees of freedom reduces this amplitude of the GW spectrum.  
The value of the new plateau due to the additional new physics (NP) degrees of freedom can be estimated as \cite{Vilenkin:2000jqa,Cui:2018rwi}
\begin{align}
    \Omega^\mathrm{NP}_\mathrm{GW}\sim \Omega_\mathrm{GW}^\mathrm{SM} \left( \frac{g_\ast^\mathrm{SM}}{g_\ast^\mathrm{SM}+\Delta g_\ast^\mathrm{NP}}  \right)^{1/3}.
\end{align}
In particular, for the MSSM, we have $\Delta g_\ast^\mathrm{SUSY}=122$, giving $\Omega^\mathrm{SUSY}_\mathrm{GW}/\Omega^\mathrm{SM}_\mathrm{GW} \approx 0.8$.
By measuring the frequency $f_\mathrm{S}$ where the GW spectrum 
starts to show deviations from the cosmic string one in standard cosmology, as well as the characteristic drop of the amplitude towards the new plateau $\Omega^\mathrm{SUSY}_\mathrm{GW}$, GW observatories can be sensitive to $\Delta g_\ast^\mathrm{SUSY}$ and $m_\mathrm{S}$.

\begin{figure}[t!]
    \centering
    \includegraphics[width=\linewidth]{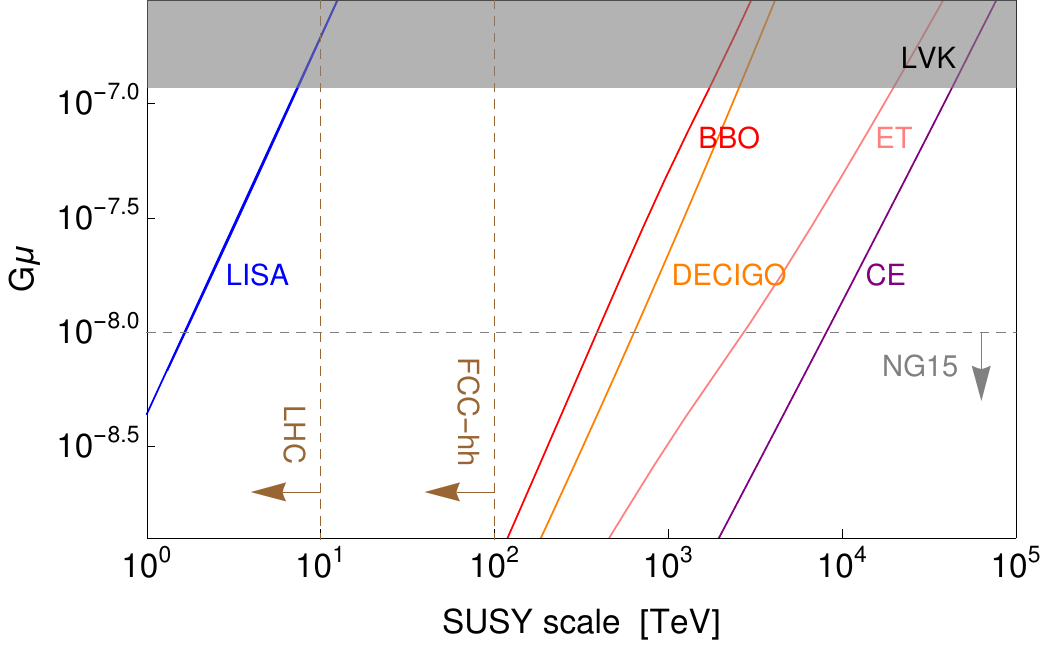}
    \caption{Lower limit on $G\mu$ for which the effect of SUSY degrees of freedom on the GW spectrum of metastable cosmic strings can be detected for various GW observatories as a function of the SUSY scale. A detectability threshold of SNR=1 for an observation time of one year is assumed. The gray shaded region represents the current LVK bound, whereas the gray dashed line indicates the 95\% credible region of NANOGrav. For comparison, the brown dashed lines show the center of mass energies for current and planned collider experiments.}
    \label{fig2}
\end{figure}

As can be seen from the above formula, $G\mu\sim 10^{-7}$, which is preferred to explain the recent PTA dataset, leads to a frequency of $f_\mathrm{S}\sim 0.06$ Hz for a 3 TeV SUSY scale, which is within the LISA sensitivity (towards the end of the band, to be specific). This is confirmed by our numerical results shown in  Fig.~\ref{fig1}. Moreover, ET and CE (BBO and DECIGO), having sensitives in the range $f\sim \mathcal{O}(10^0-10^3\mathrm{~Hz})$ ($f\sim \mathcal{O}(10^{-3}-10^1\mathrm{~Hz})$), will provide us with the fascinating possibility of probing SUSY scales from the TeV scale to much higher scales. To determine the sensitivity of these experiments we carry out a signal-to-noise ratio (SNR) analysis, where we compare the GW spectrum that includes the SUSY degrees of freedom with the one only incorporating the SM degrees of freedom. We assume standard cosmology and set the detectability threshold to SNR=1 for an observation time of one year.  See Appendix~\ref{sec:apx-b} for details of the computational procedure. The results are depicted in Fig.~\ref{fig2}. We find that high values of SNR can be obtained in multiple GW detectors. However, Advanced LVK's (HLVK's) upcoming observing run, O5, is not sensitive to SUSY degrees of freedom. As can be seen from Fig.~\ref{fig2}, LISA is sensitive to the SUSY scale in the muti-TeV range, whereas ET and CE have the potential to detect signs of SUSY up to $\mathcal{O}(10^4)$ TeV. 
Moreover, to determine how exact the number of additional degrees of freedom can be measured from the modified GW spectrum, we perform a Fisher analysis (see Appendix~\ref{sec:apx-b} for details). For $G\mu=10^{-7}$ we find the Fisher forecast measurement uncertainty of $\Delta g_\ast^\mathrm{NP}$ to be less than 10\% for $m_\mathrm{S}\leq 2\times 10^3$~TeV ($m_\mathrm{S}\leq 1\times 10^4$~TeV) at ET (CE). Also, (within the same range of $m_\mathrm{S}$) we find a Fisher forecast measurement uncertainty of around 5\% for the SUSY scale for both ET and CE. Therefore, future GW detectors can access new physics scales far beyond the reach of any planned particle collider.

 We have discussed SUSY as characteristic NP example, but of course any theory beyond the SM predicting a significant increase of particle degrees of freedom with up to $\mathcal O(10^4)$ TeV masses can be probed this way.

\section{Effects of Late Time Entropy Production}\label{sec:modulus}

So far, we have discussed the case that the universe evolves according to standard cosmology, apart from the effects of the extra degrees of freedom, for which we assumed a common mass scale $m_\mathrm{S}$. In the following, we discuss the possibility that late time entropy production, which is a characteristic possibility in SUSY models, affects the SGWB produced from metastable cosmic strings. 

For example, when SUSY is broken spontaneously in a hidden sector by the F-term of some chiral superfield, its fermionic component is eaten by the gravitino to generate its mass via the super-Higgs mechanism. But the scalar component, the sgoldstino, remains in the theory and typically has a mass around the gravitino mass. During inflation the sgoldstino is frozen at a field value displaced from its true minimum. When the temperature drops beyond a certain value, the field starts to oscillate and adds a matter-like contribution to the energy density of the universe, which at this time is radiation dominated. Since the matter-like contribution dilutes less than radiation, it can dominate the universe for some time period before it decays. When it decays, it produces entropy that dilutes also the earlier produced GWs.

Other examples of such super-weakly interacting particles, which can produce significant late time entropy (for impacts on cosmology from late decaying particles, see, e.g., Refs.~\cite{Ellis:1984eq,Endo:2006zj,Nakamura:2006uc,deCarlos:1993wie,Hasenkamp:2010if,Co:2016xti,Apers:2024ffe}), are the gravitinos themselves, moduli fields from string theory, or saxions from a SUSY solution to the strong CP problem. In all these cases it has to be ensured that the late decaying field either has negligible energy, or that it decays before Big Bang Nucleosynthesis (BBN) in order not to spoil the light element distributions (cf.~\cite{Coughlan:1983ci}),  which for only gravitationally interacting particles implies that its mass has to be $\gtrsim 10$ TeV (for studies of the compatibility of late decaying gravitinos with BBN, see, e.g.\ \cite{Ellis:1986zt,Moroi:1994rs,Moroi:1999zb,Moroi:1999zb}). Since the former case has negligible consequences for the GW spectrum, we focus on the latter.

To investigate the effects of late decaying super-weakly interacting particles on the SGWB from metastable cosmic strings~\cite{Cui:2017ufi,Cui:2018rwi,Auclair:2019wcv,Gouttenoire:2019kij,Blasi:2020wpy}, we assume an additional matter dominated epoch in the evolution of the otherwise radiation dominated universe before BBN. We assume that the matter domination phase ends at redshift $z_E$, and gives rise to a dilution factor $\mathcal D$ (i.e.\ it has started when the scale factor of the universe was a factor $\mathcal D$ smaller than at redshift $z_E$). To ensure compatibility with BBN, we consider scenarios where the late decay happens at temperature  $T_\mathrm{D}>5$ MeV. Details about the numerical implementation are given in Appendix~\ref{sec:apx-a}.

\begin{figure}[t!]
    \centering
    \includegraphics[width=\linewidth]{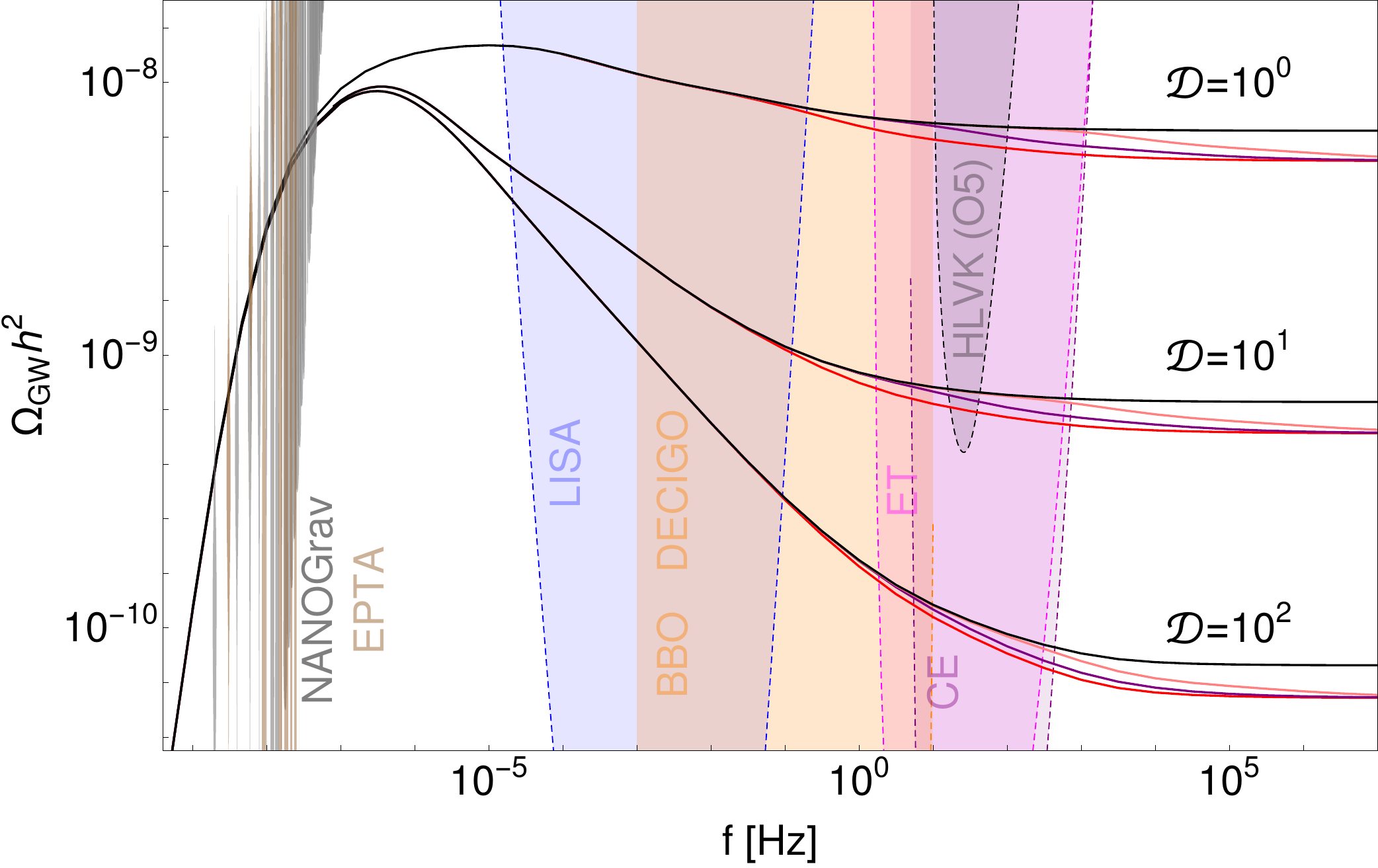}
    \caption{Effect of an additional matter domination phase on the GW spectrum of metastable cosmic strings (with $G\mu=1.17 \times 10^{-7}$ and $\kappa^{1/2}=8.115$). Spectra with dilution factors $\mathcal D=10,100$ (with $z_E=10^{10}$) are shown as well as spectra without dilution for comparison. The power-law integrated sensitivity curves of various (future) GW observatories are shown as well as the recent NANOGrav and EPTA results. The color scheme is the same as in Fig.~\ref{fig1}. }
    \label{fig3}
\end{figure}

Our numerical results are shown in Fig.~\ref{fig3}, demonstrating that the additional matter domination phase can dramatically impact the GW spectrum. The plot shows the case $z_E = 10^{10}$ with dilution factors $\mathcal D=10, 100$ and the undiluted case for comparison, and with $m_\mathrm{S} = 10^0, 10^2, 10^4~$TeV. The late decay causes a characteristic break in the GW spectrum, which can be estimated to occur at a frequency given by~\cite{Cui:2018rwi} 
\begin{align}
f_\mathrm{brk}\sim \left( \frac{8z_\mathrm{eq}}{\alpha\Gamma G\mu} \frac{t_\mathrm{eq}}{t_E}\right)^{1/2}  \frac{1}{t_0},    
\end{align}
  where $t_0$ denotes time today, $t_\mathrm{eq}$ the time of standard matter-radiation equality, $z_\mathrm{eq}$ is the redshift of standard matter-radiation equality, and $t_E\equiv t(z_E)$ is the time where the additional phase of matter domination ends.

We  emphasize that due to the drastic drop of $\Omega_\mathrm{GW}$ for frequencies above $f_\mathrm{brk}$, one can raise the value of $G\mu$ to $G\mu> 10^{-7}$ without being in conflict with the present LVK bound. Therefore, a perfect fit to the latest PTA data with $G\mu\sim 10^{-6}$ is possible already with a small dilution factor $\mathcal D\gtrsim \mathcal{O}(10^1)$. A dilution factor of order $\mathcal D\gtrsim \mathcal{O}(10^2)$  could delay the discovery of a metastable cosmic string induced SGWB beyond planned HLVK sensitivities, while still remaining in reach of future GW observatories.

It is also important to note that when the dilution factor $\mathcal D$ is large, one can argue that the GW spectrum at large frequencies $f\gg f_\mathrm{brk}$ approaches a GW spectral index $-1/3$~\cite{Blasi:2020wpy}.
For $\mathcal D \lesssim 10^2$ we find that the matter domination phase is too short for this characteristic logarithmic slope to develop. This is interesting, because (as long as the logarithmic slope has not reached $-1/3$) it enhances the possibilities to reconstruct the given model (here specified by $G\mu$, $\kappa$, $z_E$, $\mathcal D$ and $m_\mathrm{S}$) from the combined data of future GW observatories. In particular, the maximal logarithmic slope of the spectrum could be used to reconstruct $\mathcal D$. 
We furthermore note that the dilution $\mathcal D$ shifts the effects of the SUSY degrees of freedom towards lower frequencies by a factor $(\mathcal D)^{1/4}$, whereas smaller $G\mu$ would shift these effects to larger frequencies by a factor $(G\mu)^{1/2}$ (cf.\ Eq.~(\ref{eq:fS})).

Fig.~\ref{fig3} also indicates that the effect of the extra SUSY degrees of freedom could be measurable even in case of dilution from late time entropy production. As commented on in the previous paragraph, this would require to reconstruct the model parameters $G\mu$, $\kappa$, $z_E$, $\mathcal D$, e.g.\ from LISA and PTA results, such that then ET and CE can be sensitive to the extra effect from $\Delta g_\ast^\mathrm{NP}$ at $m_\mathrm{S}$. 
We leave the analysis of the reachable sensitivities in the presence of dilution from late time entropy production to a future work~\cite{prep}, where we will also discuss the impact of lower $z_E$ on the fit to the recent PTA data.

\section{Conclusions and Summary}\label{sec:con}
Leveraging recent PTA observations of a SGWB at nanohertz frequencies we examine how future GW observatories could detect variations in particle degrees of freedom indicative of SUSY. Under the assumption that the PTA signal originates from metastable cosmic strings, we demonstrate that these detectors could identify the signature doubling of degrees of freedom even for superpartner masses up to $\mathcal{O}(10^{4})$~TeV.  
We also discuss scenarios where entropy production, potentially caused e.g.\ by a late-decaying modulus field leading to an intermediate matter dominated phase, somewhat dilutes the GW spectrum. This would affect the timeline for detecting the GW background with LIGO-Virgo-KAGRA, but could nevertheless allow for detecting a sign of the extra SUSY degrees of freedom with Einstein Telescope and Cosmic Explorer.  In summary, if the explanation of the PTA results by metastable cosmic strings is confirmed, future GW observatories open up the intriguing possibility to probe extensions of the SM of elementary particles, like SUSY, up to mass scales far exceeding the reach of current and planned future colliders.

\vspace{15pt}
\section*{Acknowledgments}
We would like to thank Valerie Domcke and Kai Schmitz for fruitful discussions.

\appendix

\onecolumngrid

\section{Computing the GW spectrum of (Meta-)Stable Strings}\label{sec:apx-a}
In this Appendix we describe the used method for computing the GW spectrum of (meta-)stable cosmic strings. It is comprised of three main steps: the determination of the expansion history of the universe, the computation of the loop number density, and, last but not least, using the output of the two previous steps, the computation of the GW spectrum.

\begin{figure}
    \centering
    \includegraphics[width=0.47\textwidth]{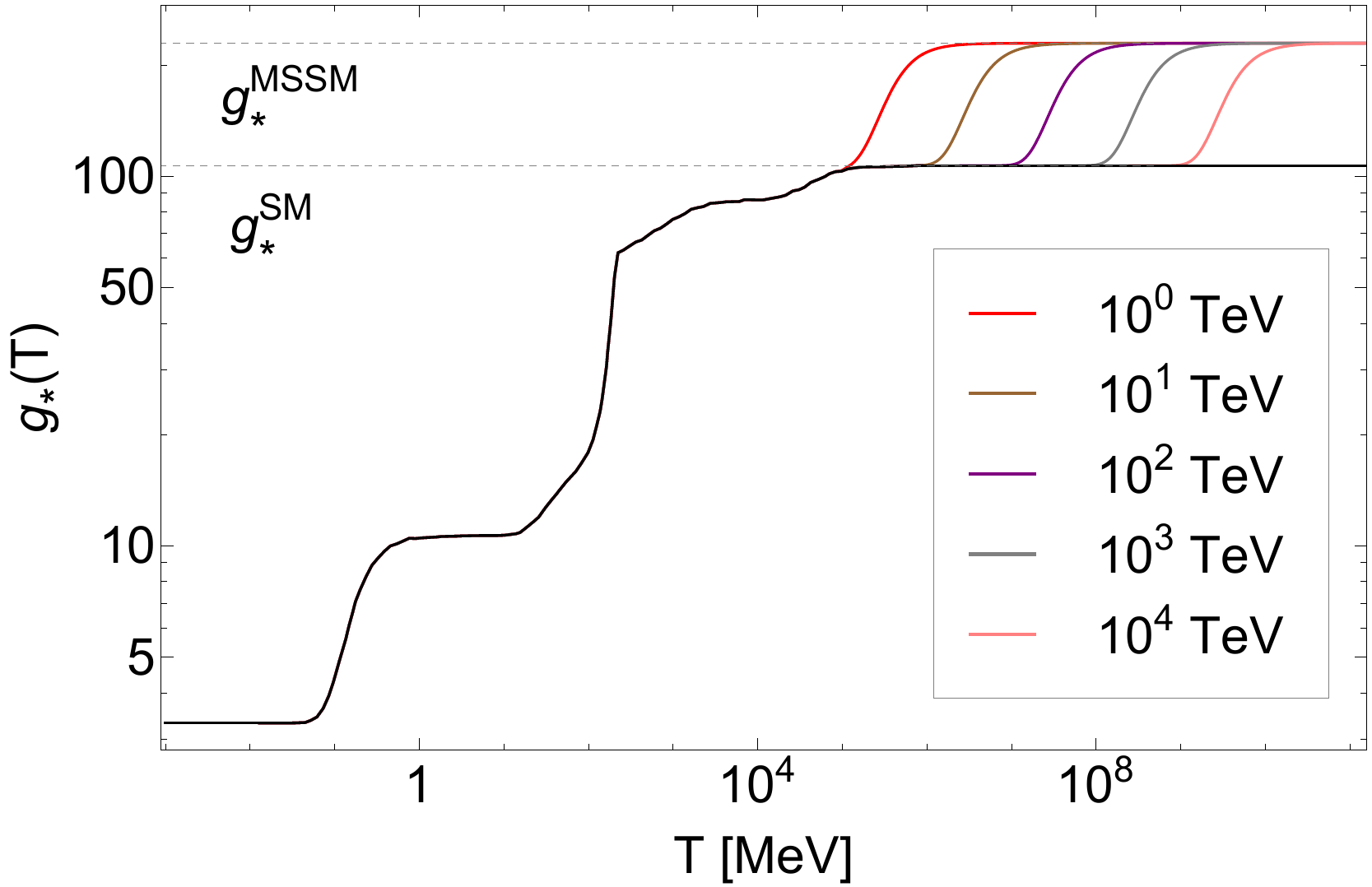}\hfill
    \includegraphics[width=0.47\textwidth]{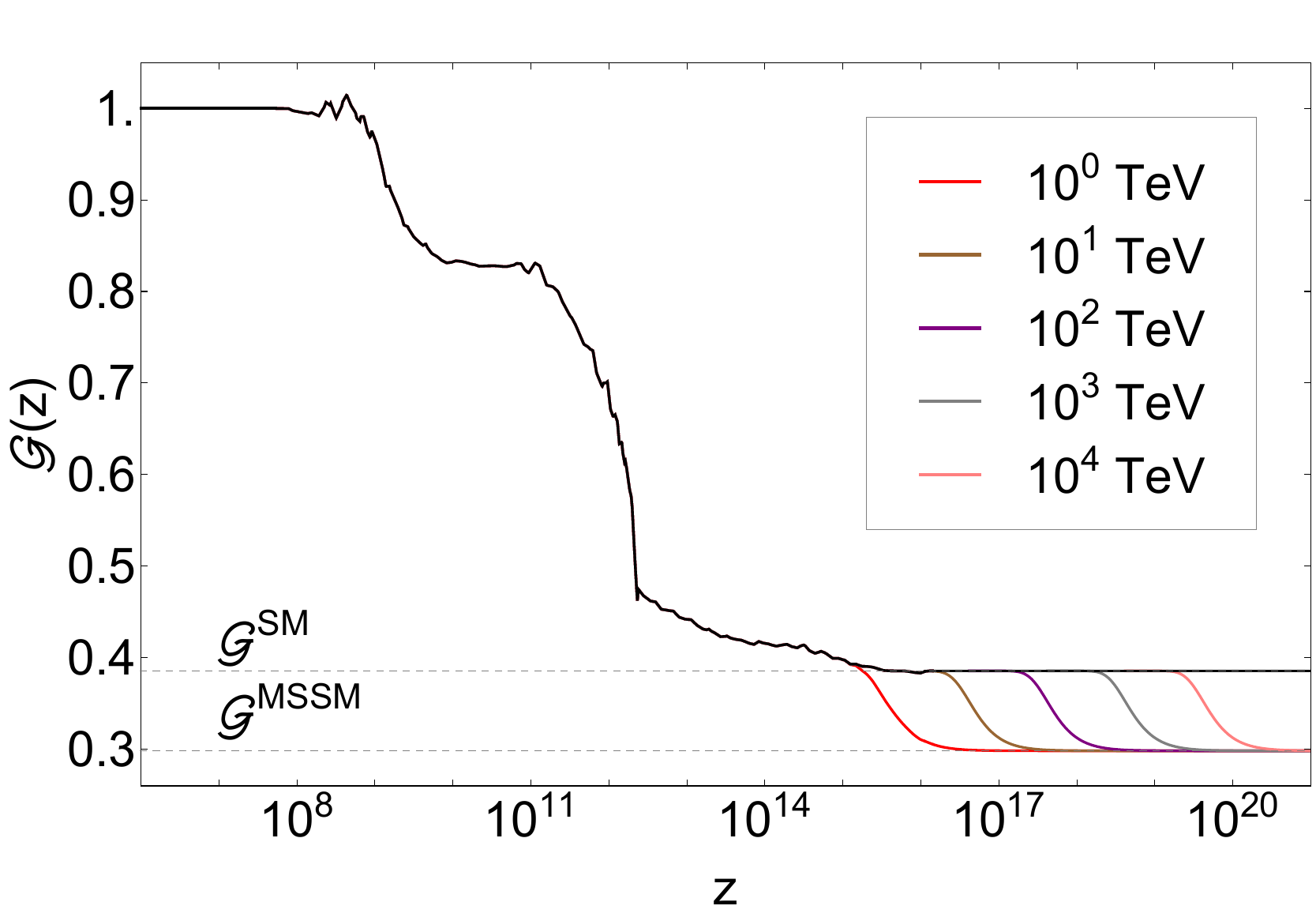}
    \caption{\textit{Left:} Plot of $g^\mathrm{MSSM}_*(T)$ for $m_\mathrm{S}=10^{0},10^{1},10^{2},10^{3},10^{4}~\mathrm{TeV}$, and $g^\mathrm{SM}_*(T)$. \textit{Right:} $\mathcal G^\mathrm{MSSM}(z)$ for $m_\mathrm{S}=10^{0},10^{1},10^{2},10^{3},10^{4}~\mathrm{TeV}$, and $\mathcal G^\mathrm{SM}(z)$. In both cases the functions where determined as described in  Appendix \ref{sec:apx-a}.}
    \label{fig:gstar}
\end{figure}

For redshift $z\leq10^6$, we take the Hubble rate to be as given by Eq.~\eqref{eq:HCDM}. In this regime $\mathcal G(z)=1$. For larger redshift, we neglect sub-leading components of the universe and only keep the dominating fluid. Hence for $z>10^6$ $H(z)$ is initially given by 
\begin{equation}
 H(z)=H_\mathrm{RD}\sqrt{\mathcal G(z)}(z+1)^2\label{eq:HRD}.
\end{equation}
with $\mathcal G(z)$ given by Eq.~\eqref{eq:Gdof} and $H_\mathrm{RD}=H_0\sqrt{\Omega_\text{rad}}$. Setting aside a possible period of matter domination, the effect of additional degrees of freedom in thermal equilibrium can then be encoded in $g_*(T)$ and $g_\mathrm{S}(T)$. In the case of the MSSM where all superpartner particles have a common mass $m_\mathrm{S}$ we get\footnote{We ignore here that neutralinos might have to be treated differently, which, however, depends on the specific model details. }
\begin{equation}
    g^\mathrm{MSSM}_*(T)=g_*^\mathrm{SM}(T)+\frac {15}{\pi^4}\bigg(32\ J_{+}\big(m_\mathrm{S}/T\big)+94\ J_{-}\big(m_\mathrm{S}/T\big)\bigg),\quad J_\pm(x)=\int_0^\infty\ d\xi \frac{\xi^2\sqrt{\xi^2+x^2}}{\exp\big(\sqrt{\xi^2+x^2}\big)\pm1}.
\end{equation}
In the regime where $g^\mathrm{SM}_*$ and $g^\mathrm{MSSM}_*$ differ substantially, the relation $g_\mathrm{S}(T)=g_*(T)$ holds, eliminating the need to compute $g^\mathrm{MSSM}_\mathrm{S}$ separately. For $g^\mathrm{SM}_{*,\mathrm{S}}(T)$ we have used the values provided by Ref.~\cite{Husdal:2016haj}. Additionally, temperature is needed as a function of redshift, which we determine by the following procedure: First, $T^\prime$ is computed at some early $z^\prime$ via 
\begin{equation}
 3 m^2_\mathrm{Pl}H^2(z)=\rho(z)=\frac{\pi^2}{30}g_*(T)T^4,
\end{equation}
taking advantage of the fact that $\mathcal G(z^\prime)=1$ for small enough $z^\prime$. Using $T^\prime$, $z^\prime$, and entropy conservation we then solve numerically for $T(z)$. In our analysis we approximate $\mathcal G(z)$ with step functions. An example has been illustrated in Fig.~\ref{fig:gstar}.

Now, we turn to the inclusion of a phase of matter domination (MD). The picture we have in mind is the following. At early times the universe is radiation dominated (RD), we denote the corresponding energy density by $\rho_\mathrm{R}$. Additionally there is a subdominant component to the total energy density, denoted by $\rho_\mathrm{M}$, which scales as matter due to, e.g., a long lived massive particle or a coherently  oscillating modulus field. Since $\rho_\mathrm{M}\propto a^{-3}$ and $\rho_\mathrm{R}\propto a^{-4}$, the matter component will eventually dominate, giving rise to a MD phase. The phase of MD should eventually end, through decay. Hence, we assume that the energy contained in $\rho_\mathrm{M}$ is transferred into $\rho_\mathrm{R}$, reheating the subdominant radiation component and making it dominant again. The start and end of the phase of MD are in principle determined by the NP model and solving the adequate equations of motion and/or Boltzmann equations. However, we opt to parameterize the phase of MD in a model independent way using the redshift at its end $z_E$ and the dilution factor $\mathcal D$, see Sec.~\ref{sec:modulus}. Going backwards in time, $H(z)$ is given by Eq.~\eqref{eq:HRD} until $z_E$, with $H_\mathrm{RD}$ and $\mathcal G$ being determined as discussed above. For $z_E\leq z<\mathcal Dz_E+\mathcal D-1\approx\mathcal Dz_E$, we take $H(z)$ to be 
\begin{equation}
    H(z)=H_\mathrm{MD}(z+1)^{3/2}.\label{eq:HMD}
\end{equation}
$H_\mathrm{MD}$ is fixed by requiring $H(z)$ to be continuous at $z_E$.  Prior to the phase of MD, $H(z)$ is again of the form of \eqref{eq:HRD}. Again, $H_\mathrm{RD}$ is fixed by continuity, the procedure to determine $\mathcal G(z)$ is the straight forward generalization of the one described above. $H(z)$ allows for computing $t(z)$ and the Hubble radius $d_H(t)$, which is required for the subsequent steps.

To determine the loop number-density $n(\ell,t)$, one has to solve the following partial differential equation \cite{Buchmuller:2021mbb}:
\begin{equation}
\big[-\Gamma\,G\mu\,\partial_\ell+\partial_t\big]n(\ell,t)=S(\ell,t)-\big(3H(t)+\Gamma_d\ell\big)n(\ell,t),\label{eq:stringkin}
\end{equation}
where $S(\ell,t)$ is the loop production function. In the above equation, relativistic effects have been neglected, cf.~\cite{Blanco-Pillado:2013qja,Buchmuller:2021mbb}.  Eq.~\eqref{eq:stringkin} has an analytic solution \cite{Buchmuller:2021mbb}:
\begin{align}\label{eq:solkin}
    n(\ell,t)&=\exp\bigg[-\int_{t_*}^td\tau\Gamma_d \mathbb L(\tau)\bigg]\Bigg\{\bigg(\frac{a(t_*)}{a(t)}\bigg)^3 n\big(\mathbb L(t_*),t_*\big)+\int^t_{t_*}d\tau\bigg(\frac{a(\tau)}{a(t)}\bigg)^3S(\mathbb L(\tau),\tau)\exp\bigg[\int_{t_*}^\tau d\xi\Gamma_d \mathbb L(\xi)\bigg]\Bigg\},
\end{align} 
with initial conditions for $n$ specified at a time $t_*$ and $\mathbb L(\tau)=\ell+\Gamma \,G\mu(t-\tau)$. As in Ref.~\cite{Buchmuller:2021mbb}, we distinguish between two regimes: early times $t<t_d=\Gamma_d^{-1/2}$ during which the exponentials can be drooped, and late times $t>t_d$ where $S=0$ justified by the fact that the cosmic strings network producing loops will have disappeared due to monopole nucleation. The solution Eq.~\eqref{eq:solkin} then simplifies to \begin{equation}
n(\ell,t)=\begin{cases} n_{S}(\ell,t),&t<t_d\\n_S\big(\ell+\Gamma G\mu(t-t_d),t_d\big)\exp\bigg[-\Gamma_d\Big(\ell(t-t_d)+\frac 12G\mu\Gamma(t-t_d)^2\Big)\bigg]\Big(\frac{a(t_d)}{a(t)}\Big)^3,&t>t_d\end{cases},\label{eq:nmaster}
\end{equation}
where 
\begin{equation}
n_S(\ell,t)=\int^t_{t_c}d\tau\bigg(\frac{a(\tau)}{a(t)}\bigg)^3S(\mathbb L(\tau),\tau).\label{eq:solnSCS}
\end{equation}
For the purpose of numerics, it is more convenient to compute
\begin{equation}
\tilde n(x,t)=\int^t_{t_c}d\tau\ a^3(\tau)S(x-\Gamma\ G\mu \tau,\tau) \label{eq:ntilde1},
\end{equation}
from which $n_S$ can easily be recovered. To incorporate different expansion backgrounds, the integral in Eq.~\eqref{eq:ntilde1} is split into several pieces such that within each piece, $H(z)$ corresponds to either Eq.~\eqref{eq:HMD} or Eq.~\eqref{eq:HRD} with $\mathcal G(z)=\mathcal G_i=\mathrm{const}.$:
\begin{equation}
\tilde n(x,t)=\sum_i\int^{t_i}_{t_{i-1}}d\tau\ a^3(\tau)S^i(x-\Gamma\ G\mu \tau,\tau).\label{eq:sum}
\end{equation}
$S^i$ is chosen according to whether the universe is in an era of MD or RD. We use the simulation results from Ref.~\cite{Blanco-Pillado:2013qja}: 
\begin{align}
    \text{MD}:\;\;\;&S(\ell,t)\approx d^{-5}_H(t)\frac{5.34}{(\ell/d_H(t))^{1.69}}\Theta\big(0.06-\ell/d_H(t)\big)\Theta\big(\ell/d_H(t)-\Gamma \,G\mu\big),\label{eq:fMD}\\
    \text{RD}:\;\;\;&S(\ell,t)\approx d^{-5}_H(t) 92.13\,\delta\big(\ell/d_H(t)-0.05\big).
\end{align}
Note that the second $\Theta$ function in Eq.~\eqref{eq:fMD} is introduced to cut off the $\ell/d_H(t)\rightarrow0$ divergence (cf.~\cite{Blanco-Pillado:2013qja}, remark after Eq.~(30)), other regulators are in principle possible. 

Integrals corresponding to a period of MD have been evaluated numerically, however in the case of RD one can easily derive an analytic expression. During one particular integration interval of Eq.~\eqref{eq:sum} the scale factor and the Hubble radius are given by
\begin{align}
    a(t)=\sqrt{2\mathcal G^{1/2}_iH_\mathrm{RD}(t-\delta t)},\qquad 
    d_H(t)=2(t-\delta t)+\Delta\sqrt{t-\delta t},
\end{align}
where $\delta t$ and $\Delta$ can be determined from the requirement that $d_H$ and $a(t)$ should be continuous functions. The result is given by
\begin{equation}
    \int_{t_{i-1}}^{t_i}d\tau\frac{a^3(\tau)}{d_H^5(\tau)}S^i(x-\Gamma\ G\mu \tau,\tau)   =c\frac{a^3(\tau_\mathrm{crit})}{g(\tau_\mathrm{crit})d_H^4(\tau_\mathrm{crit})}\Theta(\tau_\mathrm{crit}-t_{i-1})\Theta(t_i-\tau_\mathrm{crit}),\label{eq:int}
\end{equation}
where
$$
g(\tau_\mathrm{crit})=\bigg(2\alpha_0+\frac{\alpha_0\Delta}{\sqrt{\tau_\mathrm{crit}-\delta t}}+\Gamma\,G\mu\bigg)
$$ and
\begin{multline}
\tau_\mathrm{crit}=\frac12\big(2\alpha_0+\Gamma\,G\mu\big)^{-2}\bigg(\alpha_0^2\Delta^2+2(2\alpha_0+G\mu\,\Gamma)(x+2\alpha_0\delta t)-\\-\alpha_0\Delta\sqrt{8x\alpha_0+4G\mu\,x\,\Gamma+\alpha_0^2\Delta^2-8G\mu\,\alpha_0\,\Gamma\,\delta t-4(G\mu)^2\Gamma^2\delta t}\bigg).
\end{multline}
$c$ and $\alpha_0$ are numerical constants defined as $c=92.13$ and $\alpha_0=0.05$, respectively. Note that when taking the limit of an eternally radiation dominated universe, i.e.~$H_\mathrm{RD}\mathcal G^{1/2}\rightarrow H_\mathrm{RD}$, $\Delta\rightarrow0$, $t_i\rightarrow t$, $t_{i-1}\rightarrow0$ and $\delta t\rightarrow0$, and taking advantage of $\alpha_0\gg \Gamma\,G\mu$ one recovers the loop number density of \cite{Blanco-Pillado:2013qja}. It is also possible to recover the expression in Eq.~(37) of Ref.~\cite{Blanco-Pillado:2017oxo}. The benefit of our approach is that it allows for the epochs being arbitrarily small, whereas the authors of Ref.~\cite{Blanco-Pillado:2017oxo} implicitly assume $d_H\sim 2t$ which only holds asymptotically. Additionally, the true $d_H$ is typically larger than $2t$. Since $d_H$ enters with the fifth inverse power into equation Eq.~\eqref{eq:int} this can lead to an additional suppression.
Finally, the GW spectrum is evaluated via
\begin{equation}
    \Omega_\mathrm{GW}(f,t)=\frac{8\pi (G\mu)^2}{3H^2(t)}\sum_{n=1}^\infty C_n P_n,\quad  C_n=\frac{2n}{f^2}\int_{z(t)}^{z_c}\frac{dz}{H(z)(1+z)^6}n\Big(\frac{2n}{f(1+z)},t(z)\Big).
    \label{eq:OGW}
\end{equation}
We have evaluated the sum $\sum_{n=1}^\infty C_n P_n$ using the method proposed in the Appendix of Ref.~\cite{Blanco-Pillado:2017oxo} and used the $P_n$ coefficients from the same reference.
\section{Statistical Analysis }\label{sec:apx-b}

In this Appendix, we discuss the details of the data analysis method adopted in the main text.  
The GW signal emanating from a cosmic string is given in terms of the GW energy density spectrum, $\Omega_\mathrm{signal}(f)$, as a function of frequency. While the instantaneous sensitivity of a GW experiment is characterized in terms of the noise spectrum, $\Omega_\mathrm{noise}(f)$. Possessing both spectra enables an evaluation of the likelihood that the anticipated signal will be detected in an experiment. Commonly, this assessment is carried out through computing the SNR, defined as~\cite{Allen:1996vm,Allen:1997ad,Maggiore:1999vm,Romano:2016dpx}
\begin{align}
\mathrm{SNR}=  \sqrt{t_\mathrm{obs}  \int_{f_\mathrm{min}}^{f_\mathrm{max}}df  \left( \frac{\Omega_\mathrm{signal}(f)}{\Omega_\mathrm{noise}(f)} \right)^2 }. \label{SNR} 
\end{align}
Here the integration is taken over the experiment’s  accessible frequency range $f\in \left(f_\mathrm{min},f_\mathrm{max}\right)$ and its total observing time $t_\mathrm{obs}$. For an observation time of $n_T$ years, when SNR$\:\gtrsim n_T$ is obtained one can deduce that the corresponding GW experiment  possesses the capability to detect the anticipated GW signal. For finding the SNR, we compute $\Omega_\mathrm{signal}(f)$ from our theory predictions, and for $\Omega_\mathrm{noise}(f)$, we use the data provided in Ref.~\cite{Schmitz:2020syl} for various GW facilities.

In order to determine the sensitivity of the additional SUSY degrees of freedom, in Eq.~\eqref{SNR} we instead consider the following form of the GW signal~\cite{Kuroyanagi:2018csn,Caldwell:2018giq}
$\Omega_\mathrm{signal}(G\mu,f) \to \Omega_\mathrm{MSSM}(G\mu,f,m_\mathrm{S})-\Omega_\mathrm{SM}(G\mu^\prime,f)$, 
and perform a $\chi^2$-analysis to minimize the corresponding SNR as a function of $G\mu^\prime$ with fixed additional SUSY degrees of freedom, i.e., $\Delta g_\ast^\mathrm{SUSY}=122$. We utilize this statistical method to determine if the difference between the spectra, with and without the additional degrees of freedom, is significant enough relative to the noise to be detected. For experiments which will start taking data after LISA, we assume that LISA has already determined $G\mu$ and do not perform a $\chi^2$-minimization.
Moreover, a Fisher analysis allows estimating how well the number of additional degrees of freedom and their corresponding mass scale can be determined in a given GW observatory. To that end, the Fisher matrix is defined as (see e.g.\ \cite{Tegmark:1996bz})
\begin{align}
    \mathcal{F}_{ab}=t_\mathrm{obs}\int^{f_{\max}}_{f_{\min}} \frac{\frac{\partial \Omega_\mathrm{signal}(f)}{\partial\theta_a}\,\frac{\partial \Omega_\mathrm{signal}(f)}{\partial\theta_b}}{\left(\Omega_\mathrm{signal}(f)+\Omega_\mathrm{noise}(f)\right)^2}\,df,
\end{align}
where $\theta_{a,b}$ denote the parameters of the GW spectrum which are in our case chosen to be $\ln G\mu$, $\ln m_\mathrm{S}$ and $\Delta g_\ast^\mathrm{NP}$. The covariance matrix is then given as the inverse of the Fisher matrix.

\twocolumngrid
\bibliographystyle{style}
\bibliography{reference}

\end{document}